# A systematic study of spin-dependent recombination in GaAs$_{1-x}$N$_x$ as a function of nitrogen content

*A.C. Ulibarri, R. Kothari, A. Garcia, J.C. Harmand\*, S. Park, F. Cadiz, J. Peretti, and A.C.H. Rowe*

Laboratoire de Physique de la Matière Condensée, Ecole Polytechnique, CNRS, IP Paris, Palaiseau, 91128 France
*Centre de Nanosciences et de Nanotechnologies, CNRS, Université Paris-Saclay, Palaiseau, 91120 France

A systematic study of spin-dependent recombination (SDR) under steady-state optical pumping conditions in dilute nitride semiconductors as a function of nitrogen content is reported. The alloy content is determined by a fit of the photoluminescence (PL) intensity using a Roosbroeck-Shockley relation and verified by a study of the GaN-like LO$_2$ phonon peak in a Raman spectroscopy map. PL spectra taken from alloys of the form GaAs$_{1-x}$N$_x$ where $0.022 < x < 0.036$ exhibit PL intensity increases when switching from a linearly- to a circularly-polarized pump up to a factor of 5 for $x = 0.022$. This work used a 1.39 eV laser with a radius of 0.6 μm. The observed SDR ratio monotonically decreases with increasing $x$, reaching 1.5 for $x = 0.036$. Moreover, the excitation power required to obtain maximum SDR systematically increases with increasing $x$, varying from 0.6 mW for $x = 0.022$ to 15 mW for $x = 0.036$. These observations are consistent with an increase in the density of electronically active defects with increasing nitrogen content, both those responsible for the SDR as well as other, standard Shockley-Read-Hall (SRH) centers.

**1. Introduction**

The recombination dynamics of minority carriers in nonmagnetic semiconductors become spin-dependent in the presence of an SRH process occurring at a paramagnetic recombination center [1] due to the exchange interaction operating on quantum correlated electron pairs formed during the electron capture process. [2] If the spin-dependent recombination (SDR) rate via this route is faster than other SRH rates or the radiative band-to-band rate, luminescence and photo-current intensities can increase spectacularly when conduction electrons are spin-polarized. [2] The ratio of these intensities is known as the SDR ratio.

Spin polarization of conduction electrons, and the discovery of SDR, was initially achieved using resonance methods in crystalline [3] and amorphous [4] silicon. In direct gap III-V semiconductors, optical orientation with circularly polarized photons provides an efficient means to spin polarize conduction electrons, and SDR in the PL was initially observed in ternary AlGaAs alloys. [5] After a long hiatus, giant effects were observed in dilute nitrides of the form $GaAs_{1-x}N_x$ [6, 7] where typically, $x < 0.05$. Indeed, the SDR in dilute nitrides is so large that its measurement via photoconductivity [8] has catalyzed device propositions ranging from spin filters [9] to spin-photon interfaces [10] and quantum sensors acting as absolute magnetometers. [11]

In parallel with these efforts to develop device applications, significant fundamental progress in the identification of the SDR defect has been made. In particular, the spin dynamics of the paramagnetic center were studied, [12] and optically-detected magnetic resonance experiments have identified a $Ga^{2+}$ interstitial defect [13] as being responsible for SDR. The basic physical picture that describes SDR is now widely accepted – spin polarization of conduction electrons results in a dynamic polarization of paramagnetic centers that effectively quenches a *spin-dependent* SRH recombination path. This results in an overall increase in the minority-carrier lifetime and increases PL intensity via *spin-independent* radiative transitions. Here, we report a systematic study of the steady-state SDR under optical pumping conditions in dilute $GaAs_{1-x}N_x$ as a function of alloy content $x$ and excitation power. By fitting PL spectra with a two-component Roosbroeck-Shockley model [14] that accounts for the strain-induced splitting of the heavy- and light-hole bands, clear tendencies in the SDR with these external parameters become apparent.

We also validate the findings of this fitting procedure by performing Raman spectroscopy on the surface of a $GaAs_{1-x}N_x$ wafer where $x$ varies spatially. From the spectra, a GaN-like $LO_2$ phonon peak is extracted and compared to a GaAs-like second-order phonon spectrum to determine the $x$ content [15].

2. Determination of alloy content

Our samples were grown by molecular beam epitaxy. A 50 nm thick p-type silicon doped, $p = 2 \times 10^{18}$ cm$^{-3}$, $GaAs_{1-x}N_x$ layer was grown on a (001) semi-insulating GaAs substrate. The growth was terminated with a 10 nm GaAs cap layer. The nitrogen content, $x$, of one wafer, nominally $GaAs_{0.979}N_{0.021}$, was found to vary significantly across the surface. The normalized PL spectra shown in Figure 1(a) were obtained from a region of the sample several hundred microns across using a pump beam at 1.39 eV focused to a Gaussian spot with a lateral extent of σ = 0.6 µm. These spectra were taken at the points on the sample surface indicated by the colored and letter labeled dots in the brightfield image shown in Figure 1(b). The labeled dots in Figure 1(b) correspond to the spectra shown in Figure 1(a). While a majority of the wafer exhibits PL similar to the blue curve in Figure 1(a), in certain areas a clear red shift in the PL occurs, which results from band bowing due to a local increase in the nitrogen content.[17] It should also be noted that a strong decrease in PL intensity is associated with the redshift. In Figure 1(a) normalization factors are indicated for each of the spectra i.e., the red spectrum is 18.7 times less intense than the blue spectrum. The origin of this spatial variation in alloy content during growth is not clear, but it opens the way for the systematic study of SDR with $x$ on a single wafer.

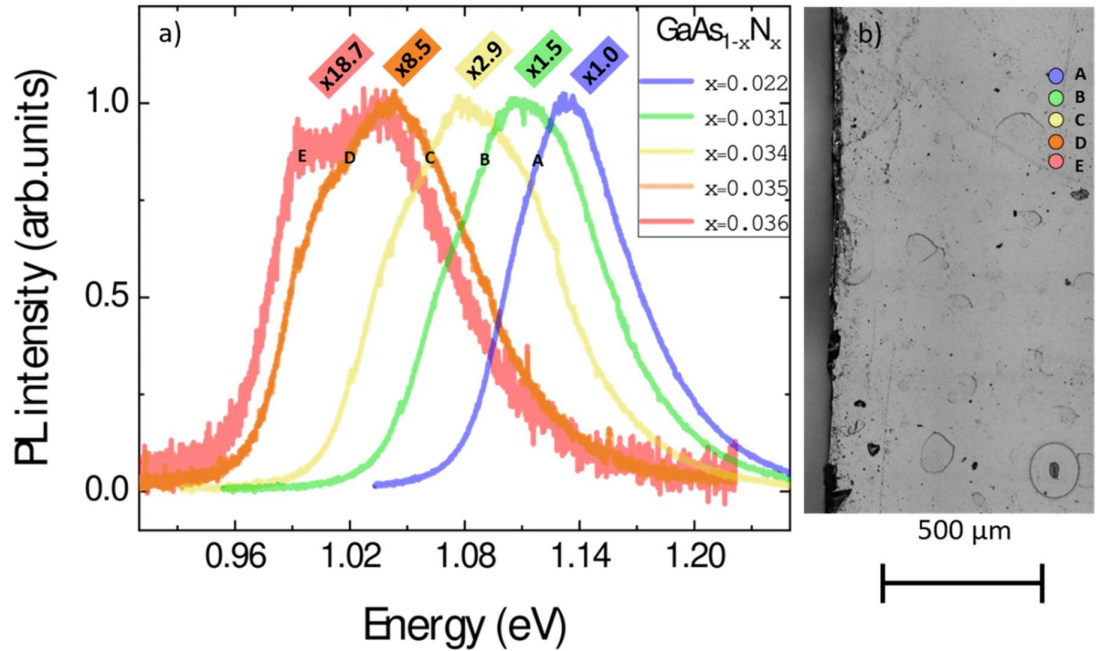

**Figure 1.** (a) Normalized PL spectra obtained at several neighboring points on a wafer whose 50 nm thick GaAs$_{1-x}$N$_x$ epilayer has a nominal nitrogen content of $x$ = 0.021. A red shift in the PL indicates a local increase in $x$ which is accompanied by a shear-stress-induced splitting of the PL peak into heavy- and light-hole components. (b) Brightfield image of the GaAs$_{1-x}$N$_x$ sample. Areas probed by PL spectroscopy are indicated by the labelled dots.

The redshift in the PL observed in Figure 1(a) arises due to a combination of an increase in nitrogen content and an increase of the in-built strain within the epilayer which, at 50 nm thick, is too thin for strain relaxation to occur. When grown on a GaAs substrate, thin dilute nitride layers experience an in-plane bi-axial tensile strain, and an out-of-plane uni-axial compressive strain, both of which increase with increasing $x$.[16] Gap changes with alloy content are then described by a band-anticrossing model [22] corrected for both the hydrostatic and shear components of the in-built strain. [18] In addition to this, the shear component of the strain also splits the heavy- and light-hole degeneracy according to

$$\delta E = 2b(\epsilon_\parallel - \epsilon_\perp), \quad (1)$$

where $\epsilon_\parallel$ is the in-plane bi-axial strain and $\epsilon_\perp$ is the out-of-plane uni-axial strain. The coefficient $b$ is the shear deformation potential of the dilute nitride.

In order to establish the nitrogen content of the five alloys with PL spectra shown in Figure 1(a), the values of the heavy- and light-hole bandgaps (and therefore the hole splitting energies) were obtained by fitting the PL spectra using a Roosbroeck-Shockley relation [14] described by

$$I(h\nu) \propto [h\nu]2\{1 - \exp[\alpha(h\nu)d]\}\exp[-h\nu/kT_c], \quad (2)$$

where $d$ is the active layer thickness, $h\nu$ is the photon energy, $\alpha(h\nu)$ is the energy-dependent absorption coefficient, and $kT_c$ is the carrier thermal energy. The absorption coefficient depends on the electronic density of states which, using Ullrich's approach, [14] switches from an Urbach tail at low energies to the 3D Bloch density-of-states at high energies.

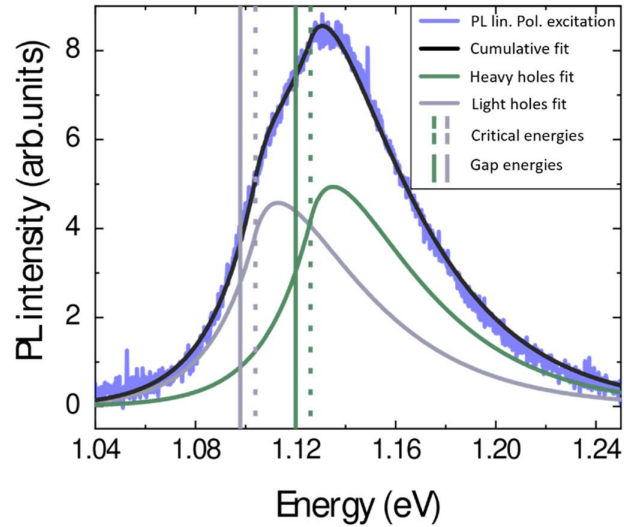

**Figure 2.** PL intensity for a GaAs$_{0.0979}$N$_{0.021}$ epilayer under linearly-polarized excitation. The fitting for light holes (gray), heavy holes (green), and the cumulative fit (black) are shown. As derived from Eq. (2), the gap values for the light and heavy holes are highlighted with solid lines and the critical energies are denoted by dotted lines.

Figure 2 shows an example of the strain-split light- and heavy-hole component peaks in gray and green, as well as their sum (in black) which was fit to the measured PL spectrum (in blue).

The so-called critical energy at which the density-of states switches from the Urbach tail to the Bloch bands is shown for each of the component peaks as a dotted, vertical line, while the energy gaps associated with each of the two transitions are shown as solid vertical lines.

Using these obtained values of the heavy- and light-hole gaps it is possible to estimate the nitrogen content, either by using the absolute values of the gaps or by using the splitting between the two. The first approach requires the use of five parameters: two empirical band coupling parameters ($V_{lh}$ and $V_{hh}$) that appear in the anti-crossing model, [17] the hydrostatic deformation potentials for the valence ($a_v$) and conduction ($a_c$) bands, and the aforementioned shear deformation potential, $b$ [18]. The second approach, on the other hand, only requires the use of a single empirical parameter, $b$, and is therefore favored here.

In the thin layer limit of relevance here, the in-plane lattice parameter of the dilute nitride is equal to the lattice parameter of the GaAs substrate, $a_\parallel = a_{GaAs} = 0.56535$ nm. The known, empirical dependence of the strain-relaxed lattice parameter, $a$, of dilute nitrides as a function of $x$ [16] was used to calculate the in-plane bi-axial strain $\epsilon_\parallel = a_\parallel/a - 1$ appearing in Eq. (1). The out-of-plane lattice constant could then be calculated using the mechanical symmetries of the face-centred cubic lattice, [18] thereby obtaining the out-of-plane uni-axial strain $\epsilon_\perp = a_\perp/a - 1$ also appearing in Eq. (1). The value of the shear deformation potential, $b$, in the dilute nitride is not clear. Some works claim $x$-dependent values as large as -3.2 eV, [19] however these claims are based on the use of Vegard's Law, which is now known not to apply in our case. [16]

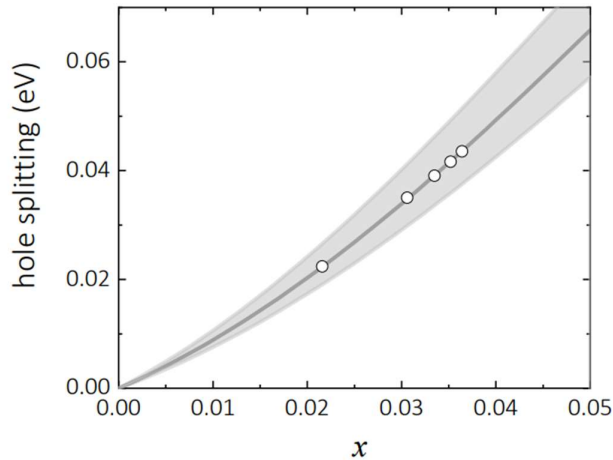

**Figure 3**. Hole splitting energies plotted as a function of alloy nitrogen content, $x$. The black line corresponds to a shear deformation potential, $b = -2$ eV in Eq. (1). The upper and lower limits of reported values for $b$ are indicated by the gray region. The nitrogen content of the five studied alloys is estimated by placing the measured hole

splittings (open circles) on the black line. The results, obtained from the (red to blue) spectra in Figure 1(a), are 0.036, 0.035, 0.034, 0.031, and 0.022. The horizontal width of the gray zone indicates the error in the estimated $x$ values.

Others have reported values ranging up to -2.4 eV for low nitrogen contents. [20] This is comparable to a value measured for GaAs of -1.7 eV. [18] There is a high degree of variability in the reported values of $b$, even for GaAs, so here a typical value of -2 eV was used to calculate the black line shown in Figure 3. Using the measured splittings obtained from the fits to the PL data, an estimate for the $x$ value can then be obtained by adjusting these points onto the black line as shown in Figure 3. From the highest to lowest nitrogen content, the five alloys studied here have nitrogen fractions $x$ equal to 0.036, 0.035, 0.034, 0.031, and 0.022 (as also noted in Figure 1(a)). To estimate the errors in this determination of alloy content, the upper edge of the gray zone in Figure 3 was calculated with $b = -2.4$ eV, and the lower edge with $b = -1.7$ eV. The width of the gray zone for a fixed hole splitting energy then gives some idea of the error (e.g., for the lowest nitrogen content alloy $x \approx 0.022 \pm 0.001$).

This fitting procedure also reveals the energy range corresponding to light emission from the Bloch density-of-states rather than the Urbach tail. In Figure 2 this corresponds to the energy range clearly above the green, dotted line but below the very high energy tail where light intensities are too low to be useful. For this particular spectrum, the heavy- and light-hole intensities to be used in the analysis of the SDR and light polarizations are therefore taken only for energies ranging from approximately 1.15 eV to 1.2 eV. A similar procedure was used on all other spectra.

### 2.1. Raman spectroscopy

The $x$ values determined from the PL spectra were verified by studying the N-related vibrational modes in a Raman scattering experiment. The spectra shown in Figure 4(b) highlight a resonance peak of Raman scattering at 470 cm$^{-1}$ which is an N-related mode. [15] These data were taken using a 1 mW off-resonance excitation at 2.62 eV focused to a Gaussian spot with a lateral extent $\sigma = 1$ μm. The N-related peak at 470 cm$^{-1}$ is associated with the GaN-related longitudinal optical (LO$_2$) phonon mode. This peak increases with intensity relative to the GaAs-related transverse optical and LO phonons whose modes are located between 500 and 580 cm$^{-1}$. Using the approach introduced by Wagner et. al., [15] a spatially resolved Raman spectroscopy map was made on the surface of the wafer, where the region where previous PL measurements have been made, indicated by the labeled dots in Figure 1(b), corresponds to the band of red pixels in Figure 4(a). Specifically, the area measured by PL spectroscopy was located on the left side of Figure 4(a) and can be identified in the brightfield images by a laser burn spot in a perfect concentric shape near the bottom left side of the image. A ring-like feature of high nitrogen, as seen by the red pixels in Figure 4(a), appears at the edge of an approximately 1 mm long region of very low, $x \approx 0.012$, nitrogen content indicated by the blue pixels near the bottom of the sample in Figure 4(a). By normalizing the GaN-related Raman peaks to those corresponding to GaAs, the nitrogen content $x$ was determined as a function of position on the sample surface. Using this approach, the highest $x$ content was found to be approximately $x = 0.037$, which agrees well with the value determined via the Roosbroeck-Shockley fit to the PL spectra discussed above.

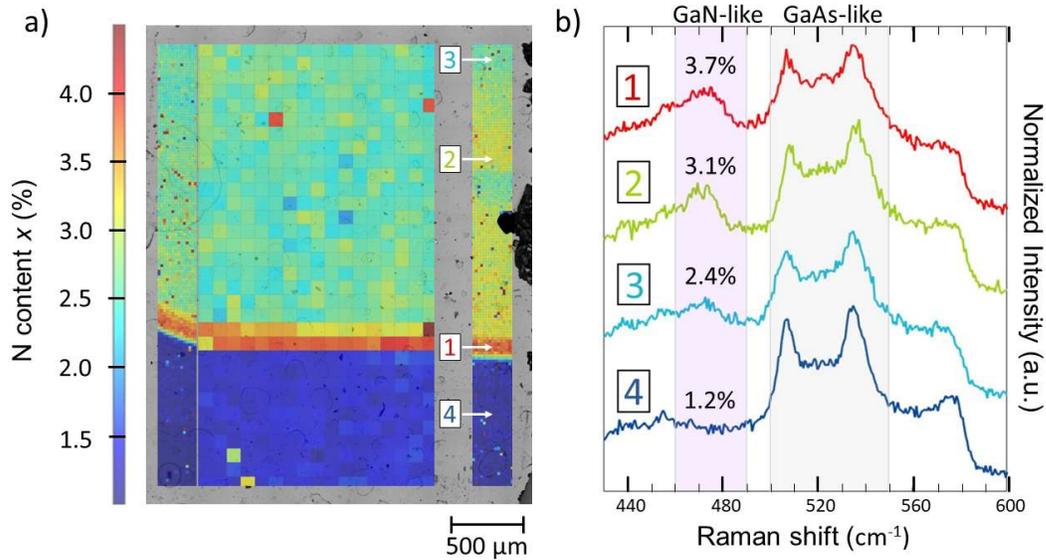

**Figure 4**. Optical microscopy and Raman spectroscopy taken from a, nominally $x=0.022$, $GaAs_{1-x}N_x$ sample. (a) Shows maps of the variation in $x$ content superimposed on an optical image of the sample. The left- and right-most maps have pixel spacing of 20 μm, while the center map has spacing of 100 μm. Pixel values represent the integrated intensity of the GaN-like $LO_2$ phonon (460 - 490 cm$^{-1}$) normalized to that of the GaAs-like second order phonon spectra (500 - 550 cm$^{-1}$) measured from the Raman spectrum at each point. [15] This spectroscopy was performed with 2.62 eV excitation at 1 mW (100x objective) with an 1800 lines/mm grating, accumulated for 30 s and averaged over 3 accumulations. Representative spectra are shown in (b) and are labeled with corresponding $x$ content.

## 3. SDR as a function of alloy content

When the sample is excited with circularly polarized laser light, a spin polarization of conduction electrons is introduced according to the optical selection rules for allowed band-to-band transitions. This in turn results in a dynamic polarization of the paramagnetic, $Ga^{2+}$, [21] centers. In steady-state the polarization of the centers is similar to that of the conduction electrons, [6] and according to the exchange interaction, this prevents recombination of the dominant spin species via these trap states. This in turn increases the PL intensity when switching from linearly- to circularly-polarized pump light. As previously described, the ratio of these two PL intensities is called the SDR and is shown in Figure 5(a) for varying excitation powers and alloy contents.

For each of the studied alloy contents, the SDR ratio shows the characteristic peak as a function of the excitation power. Maximum SDR occurs when the photo-excited conduction electron density is comparable to the defect density. A qualitative explanation for this is as follows: at low excitation powers, where the photo-electron density is small compared to the density of paramagnetic defects, only a small fraction of the defects are dynamically polarized and the SDR is small. At high excitation powers, where the photo-electron density is much higher than the paramagnetic defect density, the defects progressively transit to the doubly-occupied state, which is no longer an electronically active trap state. In this case, the SDR also drops. Quantitatively, this behavior can be captured by a coupled rate equation model for the electron, hole and trap state densities, [6] but no attempt is made here to fit this model to the presented data.

Of importance here is the observation that the power at which the peak SDR occurs increases with increasing nitrogen content, as seen in Figure 5(b). On the basis of the qualitative explanation above, this suggests that the density of paramagnetic traps increases with increasing nitrogen content. Another important observation is that the SDR monotonically decreases with increasing nitrogen content, as seen in Figure 5(b). This second observation can be qualitatively explained by an increase in other, as-yet unidentified, SRH centers through which *spin-independent* recombination rates become comparable to, or greater than, the *spin-dependent* recombination rate occurring via the $Ga^{2+}$ centers.

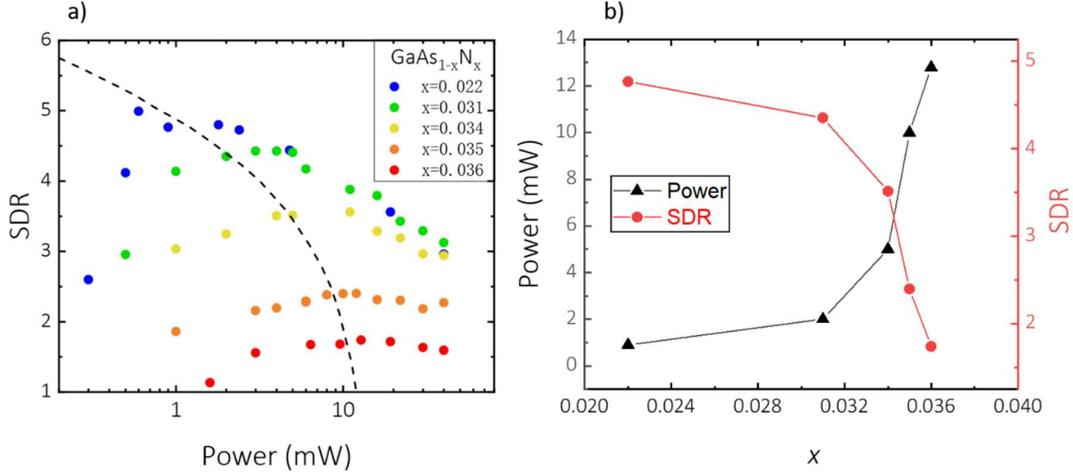

**Figure 5**. (a) Integrated SDR as a function of excitation power for different nitrogen contents. A dashed line gives a guide to the eye describing the power at which peak SDR occurs for the different $x$ values. These powers as a function of determined $x$ content are shown in (b) as black triangles and their corresponding SDR values as red dots.

## 4. Conclusions

The Roosbroeck-Shockley model relies on an assumption of electronic equilibrium and, as such, is only applicable in the strictest sense to absorption spectra. However, the excellent agreement in the values of the nitrogen content in dilute nitride alloys obtained here by comparing PL spectra with Raman spectra indicates that it can be applied to PL spectra, at least for sufficiently low excitation powers. With the variations in nitrogen content across the surface of a single wafer established in this way, we were able to perform a systematic study of the SDR as a function of nitrogen content. As the value of $x$ increased in $GaAs_{1-x}N_x$ alloys, an increase in the optical excitation power at which the peak SDR occurs was observed along with a decrease in the maximum value of the SDR. These two observations are consistent with an increase in the density of both standard SRH centers and the paramagnetic SRH centers responsible for SDR with increasing nitrogen content.

**Acknowledgments**

This work was supported by a grant from the Simons Foundation (601944, MF)

## References

[1] F. C. Rong, W. R. Buchwald, E. H. Poindexter, W. L. Warren, and D. J. Keeble, "Spin-dependent Shockley Read recombination of electrons and holes in indirect band-gap semiconductor pn junction diodes," Solid-state Electronics **34**, 835–841 (1991).

[2] D. Kaplan, I. Solomon, and N. F. Mott, "Explanation of the large spin-dependent recombination effect in semiconductors," Journal de physique lettres **39**, 51–54 (1978).

[3] D. J. Lepine, "Spin-dependent recombination on silicon surface," Physical Review B **6**, 436 (1972).

[4] I. Solomon, D. Biegelsen, and J. C. Knights, "Spin-dependent photoconductivity in n-type and p-type amorphous silicon," Solid State Communications **22**, 505–508 (1977).

[5] C. Weisbuch and G. Lampel, "Spin-dependent recombination and optical spin orientation in semiconductors," Solid-state Communications **14**, 141–144 (1974).

[6] V. K. Kalevich, E. L. Ivchenko, M. M. Afanasiev, A. Yu. Shiryaev, A. Yu. Egorov, V. M. Ustinov, B. Pal, and Y. Masumoto, "Spin-dependent recombination in GaAsN solid solutions," Journal of Experimental and Theoretical Physics Letters **82**, 455–458 (05).

[7] V. K. Kalevich, A. Yu. Shiryaev, E. L. Ivchenko, A. Yu. Egorov, L. Lombez, D. Lagarde, X. Marie, and T. Amand, "Spin-dependent electron dynamics and recombination in GaAs$_{1-x}$N$_x$ alloys at room temperature," JETP Letters **85**, 174 – 178 (2007).

[8] F. Zhao, A. Balocchi, A. Kunold, J. Carrey, H. Carrère, T. Amand, N. Ben Abdallah, J. C. Harmand, and X. Marie, "Spin-dependent photoconductivity in nonmagnetic semiconductors at room temperature," Applied Physics Letters **95**, 241104 (2009).

[9] X. J. Wang, I. A. Buyanova, F. Zhao, D. Lagarde, A. Balocchi, X. Marie, C.W. Tu, J. C. Harmand, and W. M. Chen, "Room-temperature defect-engineered spin filter based on a non-magnetic semiconductor," Nature Materials **8**, 198–202 (2009).

[10] S. Chen, Y. Huang, D. Visser, S. Anand, I. A. Buyanova, and W. M. Chen, "Room-temperature polarized spin photon interface based on a semiconductor nanodisk-in-nanopillar structure driven by few defects," Nature Communications **9**, 1–9 (2018).

[11] R. S. Joshya, H. Carrère, V. Ibarra-Sierra, J. SandovalSantana, V. K. Kalevich, E. L. Ivchenko, X. Marie, T. Amand, A. Kunold, and A. Balocchi, "Chiral photodetector based on GaAsN," Advanced Functional Materials , 2102003 (2021).

[12] V. K. Kalevich, A. Yu. Shiryaev, E. L. Ivchenko, M. M. Afanasiev, A. Yu. Egorov, V. M. Ustinov, and Y. Masumoto, "Hanle effect and spin-dependent recombination at deep centers in GaAsN," Physica B Condensed Matter **404**, 4929–4932 (2009).

[13] X. J. Wang, Y. Puttisong, C. W. Tu, A. J. Ptak, V. K. Kalevich, A. Yu. Egorov, L. Geelhaar, H. Riechert, W. M. Chen, and I. A. Buyanova, "Dominant recombination centers in Ga(In)NAs alloys: Ga interstitials," Applied Physics Letters **95**, 241904.

[14] A. Yu. Egorov, V. K. Kalevich, M. M. Afanasiev, A. Yu. Shiryaev, V. M. Ustinov, M. Ikezawa, and Y. Masumoto, "Determination of strain-induced valence-band splitting in GaAsN thin films from circularly polarized photoluminescence," Journal of Applied Physics **98**, 013539 (2005).

[15] B. Ullrich, S. R. Munshi, and G. J. Brown, "Photoluminescence analysis of p-doped GaAs using the roosbroeckshockley relation," Semiconductor Science and Technology **22**, 1174 (2007).

[16] E. P. O'Reilly, A. Lindsay, P. J. Klar, A. Polimeni, and M. Capizzi, "Trends in the electronic structure of dilute nitride alloys," Semiconductor Science and Technology **24**, 033001 (2009).

[17] W. Li, M. Pessa, and J. Likonen, "Lattice parameter in GaNAs epilayers on GaAs: Deviation from Vegard's law," Applied Physics Letters **78**, 2864–2866 (2001).

[18] Y. Zhang, A. Mascarenhas, H. P. Xin, and C. W. Tu, "Valence-band splitting and shear deformation potential of dilute $GaAs_{1-x}N_x$ alloys," Physical Review B **61**, 4433 (2000).

[19] T. Ikari, S. Fukushima, Y. Ohta, A. Fukuyama, S. D. Wu, F. Ishikawa, and M. Kondow, "Experimental study of hydrostatic and shear deformation potential in $Ga_{1-y}In_yN_xAs_{1-x}$ alloys using a piezoelectric photothermal spectroscopy," Physical Review B **77**, 125311 (2008).

[20] Wagner, J., Geppert, T., Köhler, K., Ganser, P., and Herres, N. "N-induced vibrational modes in GaAsN and GaInAsN studied by resonant Raman scattering." Journal of Applied Physics **90**, 5027–5031 (2001).

[21] Nguyen, C. T., Balocchi, A., Lagarde, D., Zhang, T. T., Carrère, H., Mazzucato, S., Barate, P., Galopin, E., Gierak, J., Bourhis, E., Harmand, J. C., Amand, T., & Marie, X. "Fabrication of an InGaAs spin filter by implantation of paramagnetic centers." Applied Physics Letters, *103*, (2013).